\documentclass[review]{elsarticle}
\usepackage{lineno,hyperref}
\modulolinenumbers[5]
\journal{Journal of Vacuum}
\usepackage{graphicx}
\usepackage{subcaption}
\usepackage{placeins}
\usepackage{multirow}
\usepackage{color}

\begin{document}

\begin{frontmatter}

\title{Flow characterization of supersonic gas jets: Experiments and Simulations}

\author[affiliation1,affiliation2]{Milaan Patel\corref{correspondingauthor}}
\cortext[correspondingauthor]{Email:milaan.patel@gmail.com}

\author[affiliation2]{Jinto Thomas}

\author[affiliation1,affiliation2]{Hem Chandra Joshi}

\address[affiliation1]{Institute For Plasma Research, Bhat, Gandhinagar,Gujarat, India, 382428}
\address[affiliation2]{Homi Bhabha National Institute, Maharashtra, India, 400094}

\begin{abstract}
In present work, we report an experimental setup that has been developed to characterize highly under-expanded helium and nitrogen jets confined in a vacuum chamber. The evolution of Zone of Silence (ZOS) is studied and found that experimentally measured ZOS for helium and nitrogen jets are in agreement with empirical relation and mathematical model. The velocity of the jet inside ZOS is measured using a developed time of fight (TOF) probe without using a skimmer. We are able to measures the velocity of the gas jet itself compared to a skimmer based method which measures the velocity of gas that is subjected to further expansion after the skimmer. Moreover, this method is free from skimmer interference and calibration involved with it. We observe that the measured velocity for helium is smaller than the terminal velocity. This can be attributed to losses which can occur due to nozzle geometry. Simulations are carried out using available DS2V code. Experimentally observed velocities of jets inside ZOS are compared with simulated values and are found to be in agreement within experimental uncertainties.
\end{abstract}


\end{frontmatter}


\section{Introduction}\label{sec:intro}

Supersonic jets have found applications in plasma diagnostics in Tokamaks TJ-II \cite{4_TJII_Atomic_beam}, TEXTOR (now decommissioned)\cite{5_kruezi_TEXTOR} and also used for fuelling \cite{10_piezo_valve,1_aditya_MBI}. Furthermore, they are also used as gas puffing sources in laser cluster interactions\cite{1_hagena_Cluster} and Z-pinch experiments\cite{1_annular_nozzle}. Requirement of high vacuum for these experiments and limitation of pumping speed has led to the use of supersonic jet sources with relatively low mass flow rates. Such mass flow rates are achieved by using a micro nozzle\cite{11_micrometer_nozzle} and low frequency operation of a gas injection system using a pulse valve\cite{2_Interferometry}. Low frequency pulses ($\sim$100Hz) are necessary to eliminate the background gas accumulation in the vessel. Moreover, this is required when used for tokamak plasma diagnostic for removing background in active spectroscopy. Also, the ability of these pulse valves to produce nearly identical pulse shapes helps to reproduce results consistently.

	Supersonic jets used in the tokamak application are generated by adiabatic expansion of a high pressure gas (1bar - 200bar) in a high vacuum ($1mbar-10^{-6}$ mbar) through a nozzle. Such an expansion results in highly under-expanded jets forming Mach cell structure after nozzle exit. The jet remains supersonic up to a certain distance inside Mach cell structure, generally referred as Zone of Silence (ZOS). Gas expanding through small size of nozzle in high vacuum results in high Knudsen number flows after nozzle exit which departs significantly from the continuum regime. The jet expands freely in vacuum and behaves like a free jet. However, the free jet has a large divergence ($>15^{o}$). When used for the fueling of a tokamak, large divergence decreases the fueling efficiency as axial velocity component decreases in the region that is far from axis of the jet limiting the penetration of the beam to the core region. Hence, such sources need to be operated in close proximity of the tokamak plasma. Moreover, close proximity also ensures fast response of the valve to enable fast chopping of gas \citep{10_piezo_valve}. Tokamak environment with large magnetic field, significant heat load, and space restriction requires such gas sources to be compact and robust. Considering these aspects it is desirable for a gas injection system to be located away from the core of the plasma. Typically such gas feed sources use capillary tube \cite{10_piezo_valve}. However, this doesnot reduce divergence and the flow is still limited by friction \cite{12_parks}. In tokamak plasma diagnostic applications, large beam divergence makes measurement of turbulent structures of edge plasma more difficult \cite{5_kruezi_TEXTOR}. Hence, a low divergence ($<2^{o}$) narrow beam is required which is typically generated using skimmers. Skimmers, because of their low divergence can be placed far ($\sim1m$) from the tokamak plasma which eliminates the need of special requirements for pulse valve. In order to make sure the extracted beam from the skimmer is supersonic, the skimmer needs to be positioned axially inside ZOS. The velocity, divergence and number density of the extracted beam depend on the axial location of the skimmer inside ZOS. Hence, it is important to characterize (particularly density and velocity) the jet upstream of the skimmer location. 

The experimental approach commonly used to measure the velocity and density of free jets is based on the time of flight (TOF) measurement using fast ionization gauge coupled with beam chopper \cite{1_Andruczyk_he} pulsed valve \cite{1_Diez_He}, mass spectrometer\cite{1_THESIS} and high power laser ionization\cite{6_skimmer_laser_photoionization}. 
In these measurements, the skimmer itself or slit is used for localized characterization of the jets by measuring the density and velocity of the extracted beam. However, the density measurements are subjected to skimmer interference\cite{6_skimmer_shock_wave} and the velocity of the extracted beam approaches the terminal velocity due to further expansion after skimmer. Hence, the measured values of velocity actually represent the characteristics of the beam, rather than the jet itself. Moreover, majority of the experiments are not aimed to measure the flow profile. Some of the studies used interferometric method to measure the density profile\cite{2_Interferometry,2_lorenz_laser}. This method provides a better characterization of the jets without perturbation. However, measurements are limited to a distance of a few nozzle exit diameters from the nozzle because the fringe shift decreases when density drops downstream of the nozzle\cite{2_lorenz_laser}. Imaging techniques have also been used in many experiments to estimate the density profiles\cite{3_Free_jets,3_belan2008,3_belan2010}. This method exploits neutral emission exited by an electron beam. The intensity of the emission is measured using a dedicated detecting system and is used to estimate the density. However, this method is sensitive to gas density, amount of ionization and sensitivity of the detection system. Hence, these measurements are usually done to characterize flow in the background pressure higher than  $10^{-2}$mbar\cite{3_belan2010}. Another TOF based method uses a microphone\cite{7_enhanced_microphone,7_spatio-temp_microphone} to estimate the velocity and density of pulsed jets. The microphone measures the differential change in capacitance of the diaphragm sensor caused by the pressure of the pulsed jet. However, this method requires high reservoir pressures (50bar) to generate dense jets to be able to be detected at a longer ($>$200mm) distance.  As far as theoretical analysis is concerned, widely used mathematical approximation taken to determine the flow behaviour is 1-D adiabatic expansion. 1-D adiabatic assumption can be used to derive the flow properties with the help of the Mach number. While inside the nozzle, Mach number can be derived from the cross sectional area ratio\cite{1_Gass_puff_Z_pinch}. However, same approach cannot be used for flows outside the nozzle exit. In order to calculate the Mach number of the flow after exit, independent information about physical parameters e.g. pressure (static or stagnantion), temperature, density (static or stagnantion)  and velocity is required. Conventionally, Mach number is determined experimentally from flow velocity. Computational fluid dynamics (CFD) tools which are based on Navier Stokes equation have limitation in simulating free jets far from the nozzle exit due to large values of local Knudsen number away from the exit. This limits the applicability of CFD simulation methods to a few nozzle-exit diameters downstream of the flow up to which the flow is in continuum region\cite{2_Interferometry}. Direct Simulation Monte Carlo (DSMC) method has been extensively used for simulating the expansion dynamics of rarefied gas flows involving low velocity gas flows \cite{14_DSMC_radiometer}, vapor expansions in vacuum \cite{15_DSMC_vapor} and study of cluster formation \cite{13_DSMC_cluster}.

Here, we would like to mention that when used with skimmers it is more desirable to have free jets to be expanded to near terminal velocity to reduce the skimmer interference down stream of the skimmer that can increase divergence of the extracted beam. Estimation of velocity is an important aspect in identifying the losses involved with most common nozzles and hence this is important from the view point of the applicability of supersonic jets in the diagnostics of edge plasma where fluctuation studies need very small divergence for finer spatial and temporal information. For efficient tokamak fueling also the beam divergence has to be smaller and for optimum gas cluster formation, beam divergence is also an important parameter. Hence a systematic study will be helpful in furthering the understanding of supersonic jets for their optimization in these areas.

Considering these aspects in the present work, we report the pressure and density profiles for helium and nitrogen jet measured experimentally, by using an inhouse developed diagnostics setup. The size of ZOS is estimated from pressure profile measured using a pressure transducer and corresponding ZOS values are compared with the mathematical and empirical models. A TOF probe based diagnostic system is developed for direct measurement of jet velocity profile inside ZOS. It is designed to minimize the reflections in the measurement device which can affect the velocity measurements. Simulations are done using DSMC code. Experimentally measured velocity profiles are compared and discussed with the simulated values. 
 
We further divide it into four sections. Section-\ref{sec:theory} describes the theory of supersonic jets and the DSMC simulation method. Experimental setup and data collection methods are described in section-\ref{sec:es}. Results are discussed in section-\ref{sec:results} and section-\ref{sec:conclusion} concludes the work.

\section{Theory and Simulation}\label{sec:theory}

\begin{figure}[ht]
\centering
\includegraphics[width=1.0\columnwidth, trim=0cm 0cm 0cm 0cm, clip=true,angle=0]{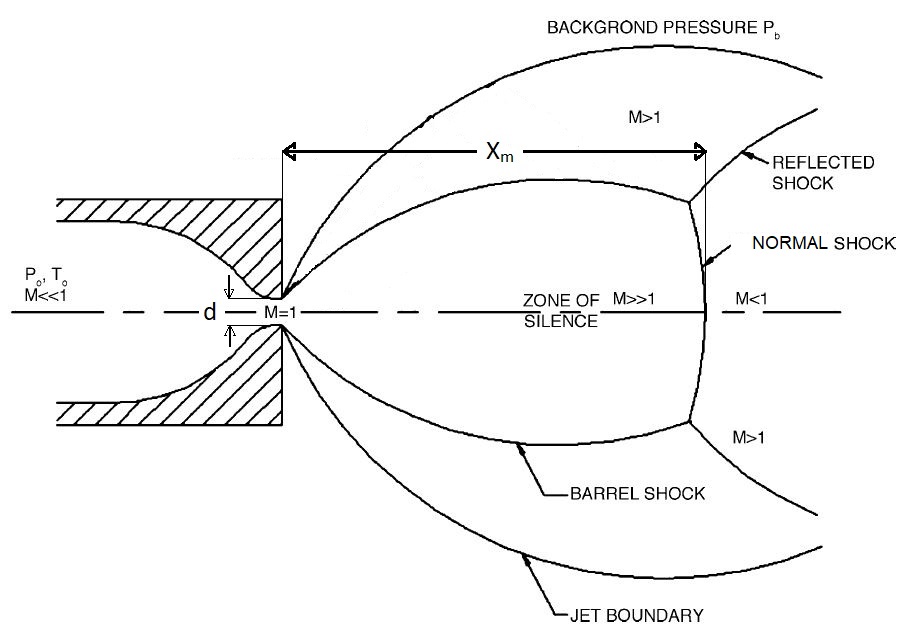}
\caption{\label{fig:ZOS} Typical Mach cell structure depicating an underexpanded supersonic jet in partial vacuum \cite{1_aditya_MBI}}
\end{figure}

A Supersonic jet can be generated using a sonic nozzle by expanding high pressure gas in low vacuum as shown in figure-\ref{fig:ZOS}. The nozzle mouth is attached to a small reservoir containing the working gas at pressure $P_0$ and stagnation temperature $T_0$. The exit of the nozzle opens towards the vacuum chamber maintained at pressure $P_b$. The pressure difference ($P_0-P_b$) drives the flow. Supersonic flow is achieved after nozzle exit if the pressure ratio $P_0/P_b$ exceeds the critical value($P_{cr}$) given by equation \ref{pr_critical}.

\begin{eqnarray}
P_{cr}=\frac{P_{0}}{P_{b}}=\left(\frac{\gamma+1}{2}\right) ^{\gamma/\gamma-1}
\label{pr_critical}
\end{eqnarray}

$P_{cr}$ is less than 2.1 for all gases. For very high pressure ratio($P_0/P_b >10$), the background pressure is significantly lower than exit pressure of nozzle, hence, the gas will continue expanding beyond the nozzle exit forming the Mach cell structure with a normal shock at the axial boundary. The distance $x_m$ between nozzle exit and normal shock is the axial length of ZOS. It depends on the pressure ratio $P_0/P_b$ and nozzle diameter($d$) given by empirical relation (equation-\ref{xm}) of Ashkenas and Sherman \cite{ash_book}. The theoretical relation given by Young using entropy balance principle \cite{5_Derivation} shows that $x_m$ is also a weak function of $\gamma$ ($C_p/C_v$) and depends on gas type (eq-\ref{xm_g}), where C($\gamma$)=0.76 for helium ($\gamma$=5/3), C($\gamma$)= 0.72 for nitrogen ($\gamma$=7/5)\cite{5_Derivation}. The experimental values of ZOS measured in the current study for helium and nitrogen jets are compared with both Ashkenas's and Young's relations.

\begin{eqnarray}
\frac{x_m}{d}=0.67 \sqrt{\frac{P_{0}}{P_b}}
\label{xm}
\end{eqnarray} 

\begin{eqnarray}
\frac{x_m}{d}=C(\gamma) \sqrt{\frac{P_{0}}{P_b}}
\label{xm_g}
\end{eqnarray} 

The supersonic expansion is theoretically formulated by 1D adiabatic expansion. Since underexpanded gas jet undergoes large radial expansion after exiting the nozzle, 1D approximation holds true only for gas flow close to the axis. Adiabatic expansion of gas inside ZOS is also an isentropic process. The stagnantion temperature of the gas remains constant throughout the flow along the axis (inside ZOS) and is equal to the reservoir temperature $T_0$. Stagnantion temperature is the measured temperature of the gas when the flow is isentropically made stationary. The kinetic energy per unit mass is given by the change in thermal energy $v^2/2=h_0 - h$, where $h_0$ and  $h$ are the specific enthalpies of the reservoir and jet respectively. Assuming that specific heat at constant pressure ($C_p$) remains constant throughout the flow, an equation relating the Mach number and stagnation to static temperature ratio can be derived (equation-\ref{t/t0}) \cite{1_Gass_puff_Z_pinch}.

\begin{eqnarray}
\frac{T_0}{T}=\left(1+\frac{\gamma-1}{2}M^{2}\right)
\label{t/t0}
\end{eqnarray} 

The velocity of the expanding gas depends on the reservoir temperature of the gas and extend to which the expanding jet cools down. Using the same energy balance equation ($v^2/2=h_0 - h$) and $C_p=\gamma R/(\gamma-1)$, equation-\ref{v} can be derived to estimate the velocity. Terminal velocity is achieved when the entire thermal energy is converted to kinetic energy. For helium and nitrogen at room temperature, the terminal velocity come out to be 1760m/s and 790m/s respectively.  

\begin{eqnarray}
v=\sqrt{\left(\frac{2\gamma}{\gamma-1}\right)R (T_0-T)}
\label{v}
\end{eqnarray} 

The simulation of gas flow is done using DSMC Method. It is based on a molecular dynamics model that simulates gas flow by simultaneously following the motions of simulated gas particles. It solves the standard equation of motions in time step one order of magnitude smaller than the mean collision time and tracks the position of the particles with each step. Each simulated gas particle represents the group of real particles. It uses a probabilistic approach to determine molecular collisions and surface interactions. The method is free from the limitations of CFD solvers to simulate flows with high local Knudsen numbers. i.e. the flows which have regions within the flow that differs significantly from the continuum. However, it can only be used for transition and free molecular flows. The DSMC method is compiled in the form of a DS2V code by G.Bird\cite{0_bird2005} and is available as an open-access code. DSMC code in conjunction with experiments was used by Even. et al.\cite{0_even_SMBI} to characterise the shaped nozzle for maximum centreline intensity. A more detailed study for various opening angles of the nozzle on beam densities was carried out by Luria. et al.\cite{0_kluria_SMBI}. Both studies show that maximum beam densities are produced by a conical nozzle with $20^o$ opening half-angle. However, the primary assumption in both simulation studies is that the jet is expanding in vacuum. Here, we have assumed that gas is expanding in the background of finite pressure. This condition is computationally more difficult to simulate over a very large domain as the program has to take into account the presence of background gas and has to scale the simulated particles accordingly. Moreover, in the present study we compare the simulated velocity directly with simulated values for a significantly large flow domain compared to above cited studies where gas density compared.  

\begin{figure}[ht]
\centering
\includegraphics[width=1.0\columnwidth, trim=0cm 0cm 0cm 0cm, clip=true,angle=0]{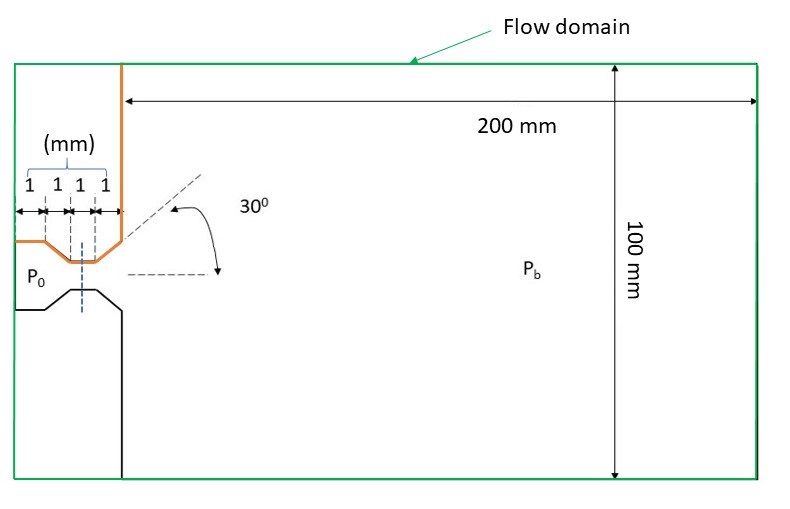}
\caption{\label{fig:flowdomain} Flow domain for DSMC simulation}
\end{figure}

The flow domain used for simulation is a rectangular box as shown in figure \ref{fig:flowdomain} (green box). Boundary conditions in DSMC simulation are applied at the ends of this rectangular box. The left is inlet and the right is exit and the top and bottom are the adiabatic walls. The inlet and exit are set as constant number density boundaries corresponding to reservoir pressure $P_0$ and background pressure $P_b$, respectively. Top and bottom walls are considered as spatially reflecting boundaries, which are analogous to adiabatic walls. The nozzle geometry is defined by two adiabatic walls inside the rectangular flow domain. These walls define the shape of the nozzle geometry as that used in present experiment setup. The exit cone half-angle is set to $30^0$ which is taken as that of the pulse valve. The flow domain is divided into upstream and downstream regions by a partition line (blue dotted). Initially, both upstream and downstream regions are filled with gas at the number densities corresponding to inlet and exit boundary conditions. As the simulation progresses, the gas from the upstream chamber flows into the downstream chamber. The program calculates the simulated particle flux leaving the upstream region through the interfacing boundary and adds the same influx at the inlet of the flow domain. During the initial stages of simulation, particle out-flux from the exit is not the same as the influx through the inlet. This increases the number of particles inside the flow domain which results in the initial unsteady nature in the DSMC simulation. As simulation progresses, the number of particles inside the flow domain saturates and eventually steady-state is achieved. The simulation results of steady-state flow are considered for comparison. 

The supersonic jet expansion system has high-density gradients spanning from high pressure at the reservoir $P_0$ to much lower pressure at outer expansion regions $P_b$. DSMC simulation scales the number of simulated particles across the whole pressure range linearly. It has to maintain a smaller number of particles for better computational efficiency for the high-pressure region while maintaining a sufficient number of particles in the outer low-pressure expansion region. This requires a large computation time for flows with high pressure variation in a large simulated area. Hence in the present work, the pressure ratio for DSMC simulation is capped to $10^4$ considering the computation time. At higher pressure ratio, the computation time is significantly high to achieve a steady-state.

\section{Experimental Setup}\label{sec:es}

\begin{figure}[ht]
\centering
\includegraphics[width=0.90\columnwidth, trim=0cm 0cm 0cm 0cm, clip=true,angle=0]{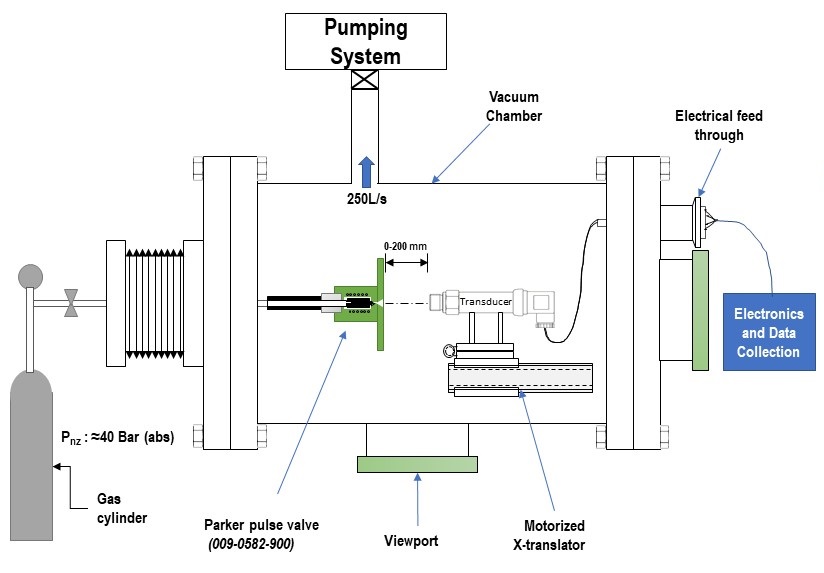}
\includegraphics[width=0.6\columnwidth, trim=0cm 0cm 0cm 0cm, clip=true,angle=0]{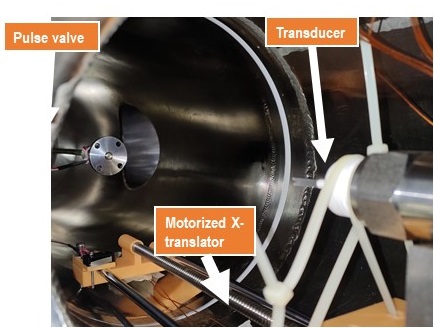}
\caption{\label{fig:expsetup} A schematic diagram of  experimental setup. The image shows the pressure transducer mounted on a X-translator, which is replaced by TOF probe (schematic shown in fig\ref{fig:tof_construction}) for TOF experiments.}
\end{figure}

The experimental setup is shown in figure-\ref{fig:expsetup}. Two different types of experiments are performed. In the first experiment, axial pressure distribution and the length of ZOS are measured by a pressure transducer (BD Sensor: DMP 320), at background, $P_{b}$, in the range of  1 mbar to 5 mbar. Accuracy of pressure transducer is $\pm$ 1\% at 10mbar. In the second experiment, axial velocity measurements are done by TOF probe in high vacuum range of the order $10^{-4}$ mbar to $10^{-6}$ mbar. The gas injection source used in both experiments is an electromagnetic pulse valve. The pulse valve is operated at pressure (reservoir pressure) $P_{0}$, in the range of 30 bar to 50 bar. All the experiments are performed for the same background gas as that of injected gas. A more detailed description of the individual system is given in the subsections below.

\subsection{Vacuum vessel}\label{subsec:vv}

The vacuum vessel used for the experiment is a cylindrical chamber, 870 mm long having an internal diameter of 200 mm. The total volume of the vessel is approximately 30 liters. Primary pumping of the vacuum vessel for the transducer experiment is done by a dry pump, BOC Edwards XDS 10, with a pumping speed of 10L/s. For TOF experiments, the pumping is done by turbo-molecular pump (TMP), Pfeiffer Hi-pace 300, with pumping speed of 250l/s and a dry pump used as backing pump. Background pressure during transducer experiments is measured with accuracy of 5\% (in range of $10^{-3}$ to $100$ mbar) by a micro-Pirani gauge (MKS Instruments) calibrated for helium and nitrogen gases. For TOF-experiments a Pfeiffer PKR251 gauge (accuracy 30\% in range $10^{-8}$ to $100$ mbar) is used to monitor background pressure. Here we would like to mention that PKR251 gauge is used only to monitor the pressure and its accuracy is not reflected in the measurements reported in the study. The gas source is mounted at one end of the chamber and electrical connections are routed through an electronic feed-through mounted at the opposite end. TMP and pressure gauges are mounted close to the pulse valve to maximize throughput and conductance, respectively. The experimentally measured ultimate vacuum attained in the vacuum vessel is $5\times10^{-7}$ mbar. 

\FloatBarrier

\subsection{Pulse valve}\label{subsec:pv}

A commercially available solenoid pulse valve: Parker, part no: 009-0181-900, with orifice size 0.8 mm is used as a gas source for supersonic jet. The pulse valve consists of a small cavity connected to a high-pressure inlet by a standard 1/4" A-Lock connector and the cavity acts as a reservoir. The end of the cavity is connected to a small orifice of diameter 0.8 mm, which opens outside the valve. The cavity is isolated from the orifice via PTFE poppet. The poppet is spring-loaded at the orifice which keeps the valve closed when not in operation. During operation, the poppet is actuated by a solenoid and the orifice opens to the cavity. There is a small conical section where the poppet locks it to the orifice. This acts as a converging section when the valve is open. The orifice opens outside the nozzle via exit cone of half-angle of $30^o$ (measured) which acts as a diverging section. 

The pulse valve is mounted inside the vacuum chamber and high-pressure connection is given through a UHV gas feed-through. The poppet sealing of the pulse valve has a rated leak rate of $10^{-7}$ (mbar l $s^{-1}$) of helium which is significantly small to affect the experiments. The valve is operated in pulse mode using dedicated IOTA-One controller which uses square pulse to trigger the valve. The opening duration can be adjusted by changing the pulse width of the input pulse. The controller has a trigger output channel that is used to synchronize the trigger pulse with the acquisition of data of transducer/TOF probe using an oscilloscope. The pulse valve is mounted on a manually operated UHV compatible vacuum bellow, which can provide axial movement of 100 mm. A bellow is used for the initial positioning of the pulse valve relative to transducer/TOF-probe, for measurements, however, a motorized translation stage is used. The translation stage has a total travel length of 230 mm and a positioning accuracy better than 1 mm in its entire travel range.

\subsection{Pressure Transducer}\label{subsec:pt}

The pressure transducer is mounted axially facing the pulse valve and is coupled with a small pitot tube of length 2 cm, internal diameter of 2 mm, and wall thickness of 0.5 mm. The transducer is mounted on a translation stage and pressure is recorded along the flow axis with a spatial resolution of 2 mm. The distance between the exit of the pulse valve and the tip of the pitot tube is considered as the separation $D$. For each $D$, the measurements are averaged for 5 pulses. The gas pulse time (valve duration) for all transducer experiments is kept 5 ms. It is necessary to keep the valve opening duration as small as possible to avoid elevating the background pressure significantly during valve opening. The opening time of 5 ms has been chosen considering the  rise and fall times of 1ms for the transducer. Here, we have assume that the measured pressure values correspond to the pressure at the tip of the pitot tube.

\begin{table}
\centering
\begin{tabular}{|c|c|c|c|c|c|}
  \hline
  & $P_0$(bar) & $P_b$(mbar) & $P_0/P_b$($\times10^4$) & $\Delta P_0$mbar(He) & $\Delta P_0$mbar($N_2$) \\
  \hline
  \multirow{5}{*}{Set-1}& 40 & 1 & $4$   & & \\ \cline{2-4}
                        & 40 & 2 & $2$   & & \\ \cline{2-4}
                        & 40 & 3 & $1.33$ & 0.20 & 0.08\\ \cline{2-4}
                        & 40 & 4 & $1$   & & \\ \cline{2-4}
                        & 40 & 5 & $0.8$ & & \\ \hline
  \multirow{3}{*}{Set-2}& 30 & 3 & $1$   & 0.18 & 0.06\\ \cline{2-6}
                        & 40 & 4 & $1$   & 0.20 & 0.08\\ \cline{2-6}
                        & 50 & 5 & $1$   & 0.24 & 0.10\\ \hline
\end{tabular} 
\caption{\label{table:trans} Operating conditions for transducer experiments: $\Delta$P represents the pressure rise after pulse.}
\end{table}

Measurements are recorded for both helium and nitrogen jets operating at different reservoir and background pressures. Table-\ref{table:trans} shows the combinations of background and reservoir pressures used in the measurement, which are divided into two sets. The first set is used to determine the size of ZOS for different pressure ratios, by changing background pressure ($P_b$), and keeping the reservoir pressure ($P_0$) constant. The pressure profiles are then used to estimate the length of ZOS. The second set of experiments is carried out for different $P_0$ and $P_b$, such that the pressure ratio remains constant. This is to confirm that the length of ZOS depends on pressure ratio and not on the background pressure. The rise in background pressure after gas injection is about 0.20 mbar for helium and 0.08 for nitrogen. During the experiment, $P_0$ was maintained within $\pm$1bar, while $P_b$ was maintained within $\pm$0.01 mbar. The variation in measured $\Delta P_0$ was $\pm$0.02 mbar and $\pm$0.01 mbar for helium and nitrogen respectively.

\begin{figure}[ht]
\centering
\includegraphics[width=0.9\columnwidth, trim=0cm 0cm 0cm 0cm, clip=true,angle=0]{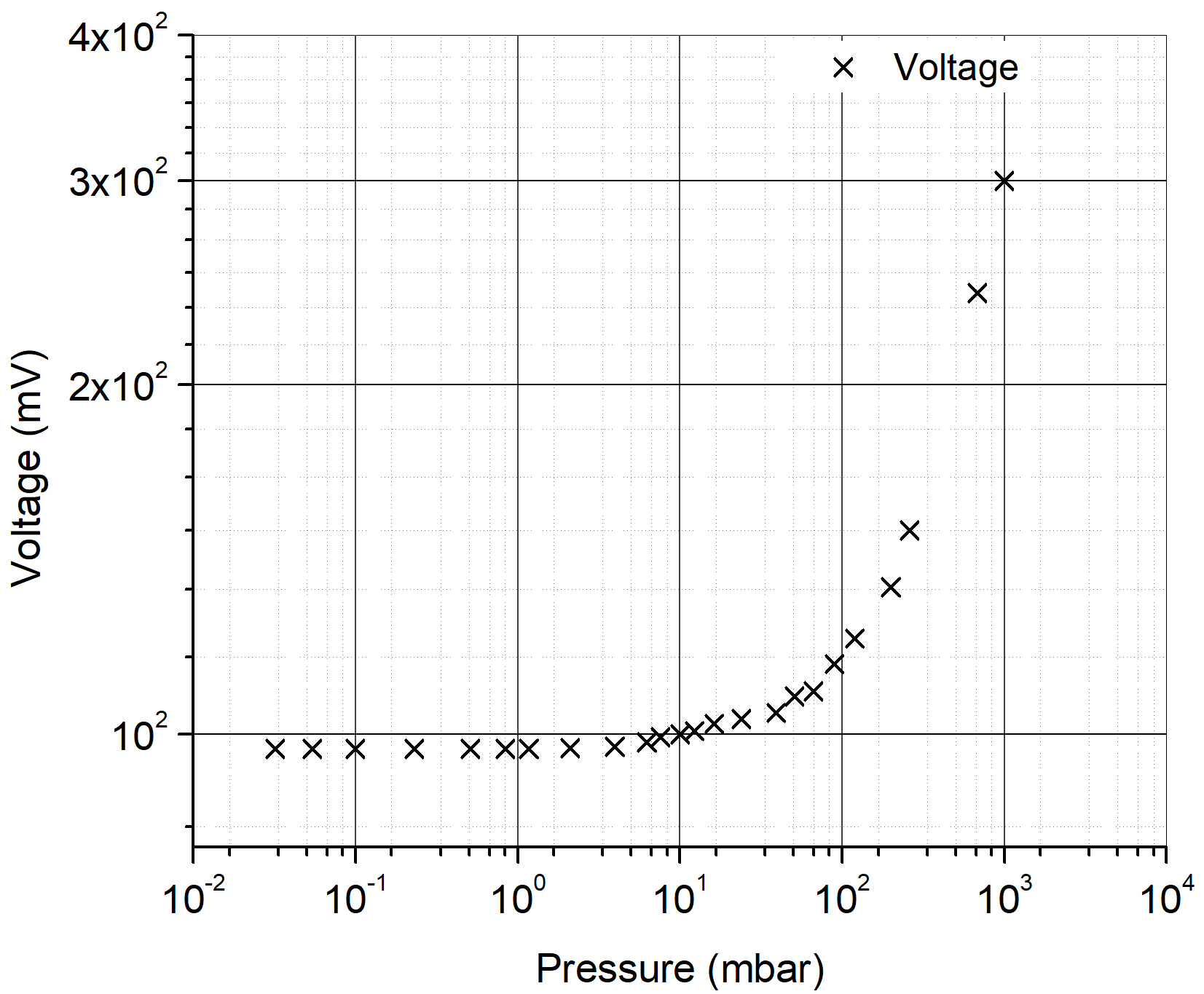}
\caption{\label{fig:transducer_PV_exp} Experimentally measured transducer output voltage with pressure}
\end{figure}

The pressure transducer is calibrated experimentally, using Pirani and Bourdon gauges. The calibration curve is shown in figure-\ref{fig:transducer_PV_exp}. From the calibration curve it is evident that the transducer is sensitive to pressures above 10mbar. As mentioned latter (results section), ZOS length is estimated by locating the peak pressure of normal shock. For background pressure higher than 1 mbar, the normal shock structure has got pressure higher than 10 mbar and is detectable by the transducer. However, it is observed that for background pressure lower than 1 mbar the normal shock is diffused and below the detection limit for the transducer. Hence 1 mbar is considered as the limit of background pressure for transducer experiments

\subsection{Time of Flight(TOF) Probe}\label{subsec:tof}

Measurement to characterize jets expanding in low vacuum should be able to detect temporal variations at low pressures that are typically present in the expansion regions of a jet. In the developed TOF probe, the gas jet is ionized by thermal electrons accelerated using an electric field. The ions are collected using a small collector wire placed inside the flow which records current proportional to the number of ions. This ion current is used to detect and characterise the gas jet. Ion current is recorded for different locations of the probe in front of the nozzle along the axial direction and the difference in travel time between two the locations is used to estimate the velocity. As the probe has a good temporal resolution (microseconds), it can be used to measure the velocity by recording travel time for differences in path lengths up to 10mm. However, for density measurement, information about the extent of ionization is required, which cannot be reliably determined. Hence, the TOF probe is only used to measure the velocity of a gas jet.

The experiments using the TOF probe are aimed at measuring the velocity profiles within the ZOS at background pressure corresponding to typical tokamak operating conditions. Hence, the TOF measurements are conducted at a background pressure of $P_b$=$5\times10^{-5}$mbar with reservoir pressure of $P_0$=40 bar. Such high-pressure ratio results in ZOS which is physically beyond the size of the test vessel. Hence the measured experimental value of flow velocity should be inside the ZOS and expected to be supersonic. Short valve opening duration ($200\mu s$) along with high pumping speed ($250l/s$) via TMP are used to minimize the rise in background pressure during gas injection.

\begin{figure}[ht]
\centering
\includegraphics[width=1.0\columnwidth, trim=0cm 0cm 0cm 0cm, clip=true,angle=0]{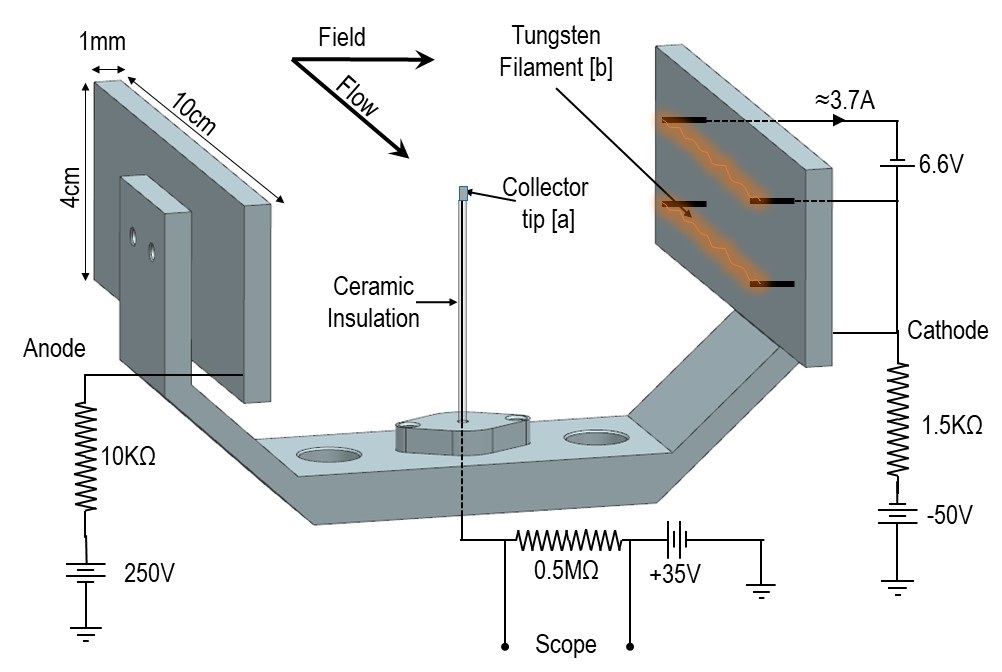}
\caption{\label{fig:tof_construction} TOF Probe schematic and electrical connections. [a] copper wire radius:2mm length:6mm, [b] pure tungsten filament 0.1mm diameter, 2 filaments are connected in parallel}
\end{figure}

Figure~\ref{fig:tof_construction} shows the TOF probe used for velocity measurement. It consists of stainless steel (SS304) 1 mm thick anode and cathode plates of 8cm$\times$4cm size separated by 10 cm with a small ion collector in the middle. Larger separation between electrodes is desirable as it minimizes their interference with the jet. However, separation can only be increased up to a certain extent without distorting the uniform electric field between them. 10 cm distance is found to be optimum for given size of electrodes and vacuum vessel of 200 mm diameter. The cathode plate has two pure tungsten filaments of diameter 0.1 mm and length 7 cm respectively which act as source of thermionic emission. Moreover, thin filament decreases radiant heat load to chamber components, especially pulse valve. To avoid heating, the experiments were performed in small sets lasting 10 minutes. The filament is biased to the cathode potential which is maintained at -50V and the anode is biased to +250V. The presence of two filaments helps to have more uniform distribution of emitted electrons over the cathode surface. The ion collector is a copper wire of 2mm diameter, which is insulated by ceramic tube except at the tip of height 6mm. It is biased to +35V and is mounted such that it is positioned along the axis of the jet. 

During operation, thermal electrons emitted from the heated filament are accelerated along the electric field oriented perpendicular to the flow direction and are collected at the anode. The collected thermal electrons along with the electrons of ionized gas result in anode current. The accelerated electrons cause ionization of the gas in the entire volume enveloped by electrodes. Most of the ions are swept away by the electric field and are collected at the cathode resulting in cathode current. The ion collector only collects the ions in a small volume close to the collector which results in collector current. The collector current is the TOF probe signal used to determine the velocity of a gas jet. Hence the axial resolution of the TOF probe depends on the collection region of the ion collector. 

\begin{figure}[ht]
\centering
\includegraphics[width=0.8\columnwidth, trim=0cm 0cm 0cm 0cm, clip=true,angle=0]{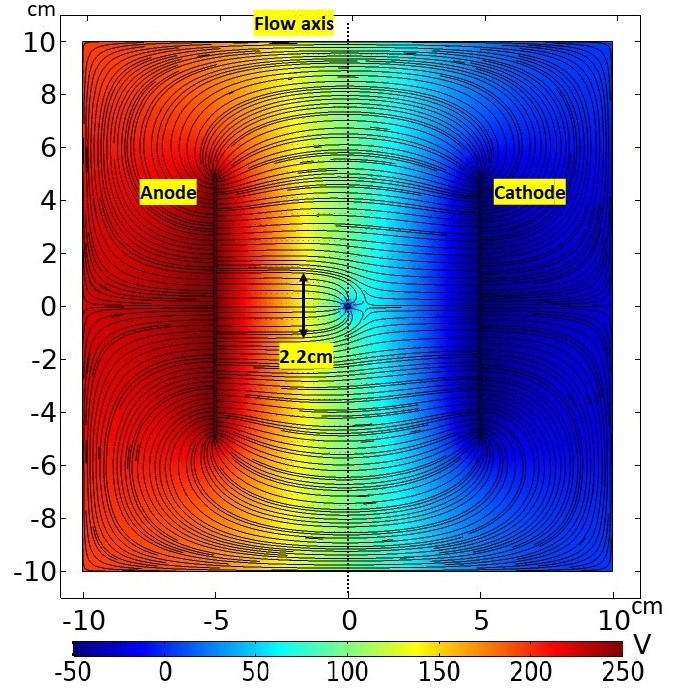}
\caption{\label{fig:e_axial} Electric field lines and potential distribution in TOF probe along axis simulated using COMSOL. The ions are collected by ion collector with volume of length in 2.2cm along axis, which indicates the axial resolution of TOF probe.}
\end{figure}

Figure-\ref{fig:e_axial} shows simulated electric field distribution of TOF probe using AC/DC module of COMSOL. The field lines along with potential distribution are plotted for the plane having axis of the flow. The field distribution shows that the axial length of collection volume is about 2.2 cm, which can be considered as typical axial resolution of the TOF probe. Resolution can be increased by reducing the size of the collection region and can be achieved by increasing the collector potential. However, improving the probe resolution does not affect the measured velocity because the velocity is measured by taking the difference of peak times, which remains same. Moreover, improving resolution reduces signal strength. Hence, probe potential of 35V was found optimum considering the trade-off between resolution and signal strength. 

\section{Results}\label{sec:results}

As mentioned, a transducer is used to measure the pressure distribution and density distribution along the axis of the jet. The ZOS length is estimated by measuring increase in pressure at normal shock. The change in ZOS for different $P_0/P_b$ ratios for helium and nitrogen jets are estimated and results are compared with empirical relation and theoretical model. It can be mentioned here that the velocity distribution cannot be derived from the transducer as the response of the transducer is too slow to detect shift in pulse with distance. Hence, as will be discussed latter, TOF probe is used to estimate the velocity distribution for helium and nitrogen jets. 

\subsection{Estimation of ZOS by Pressure Transducer}\label{subsec:results_Trans}

\begin{figure}[ht]
\centering
\includegraphics[width=0.9\columnwidth, trim=0cm 0cm 0cm 0cm, clip=true,angle=0]{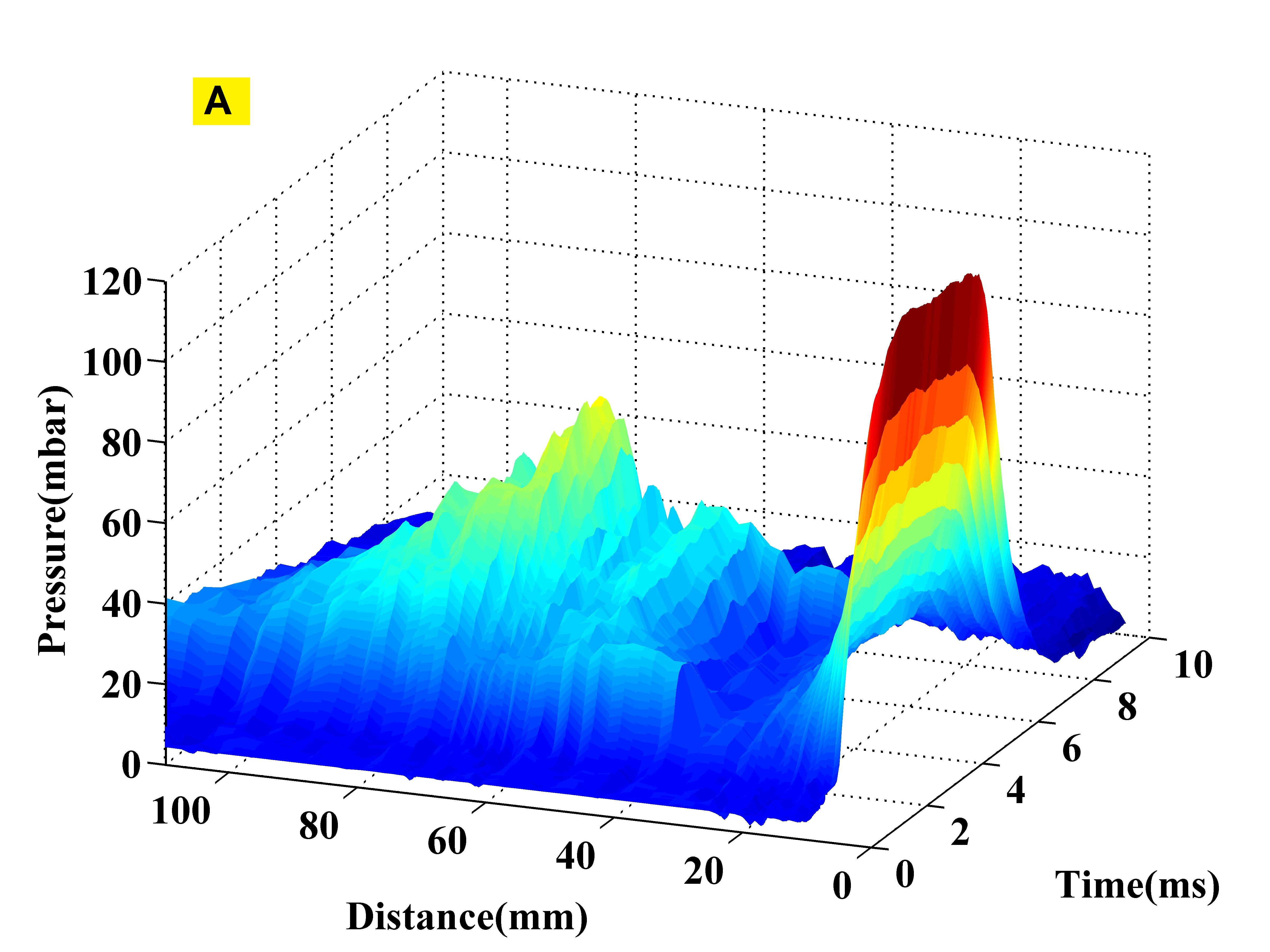}
\includegraphics[width=0.9\columnwidth, trim=0cm 0cm 0cm 0cm, clip=true,angle=0]{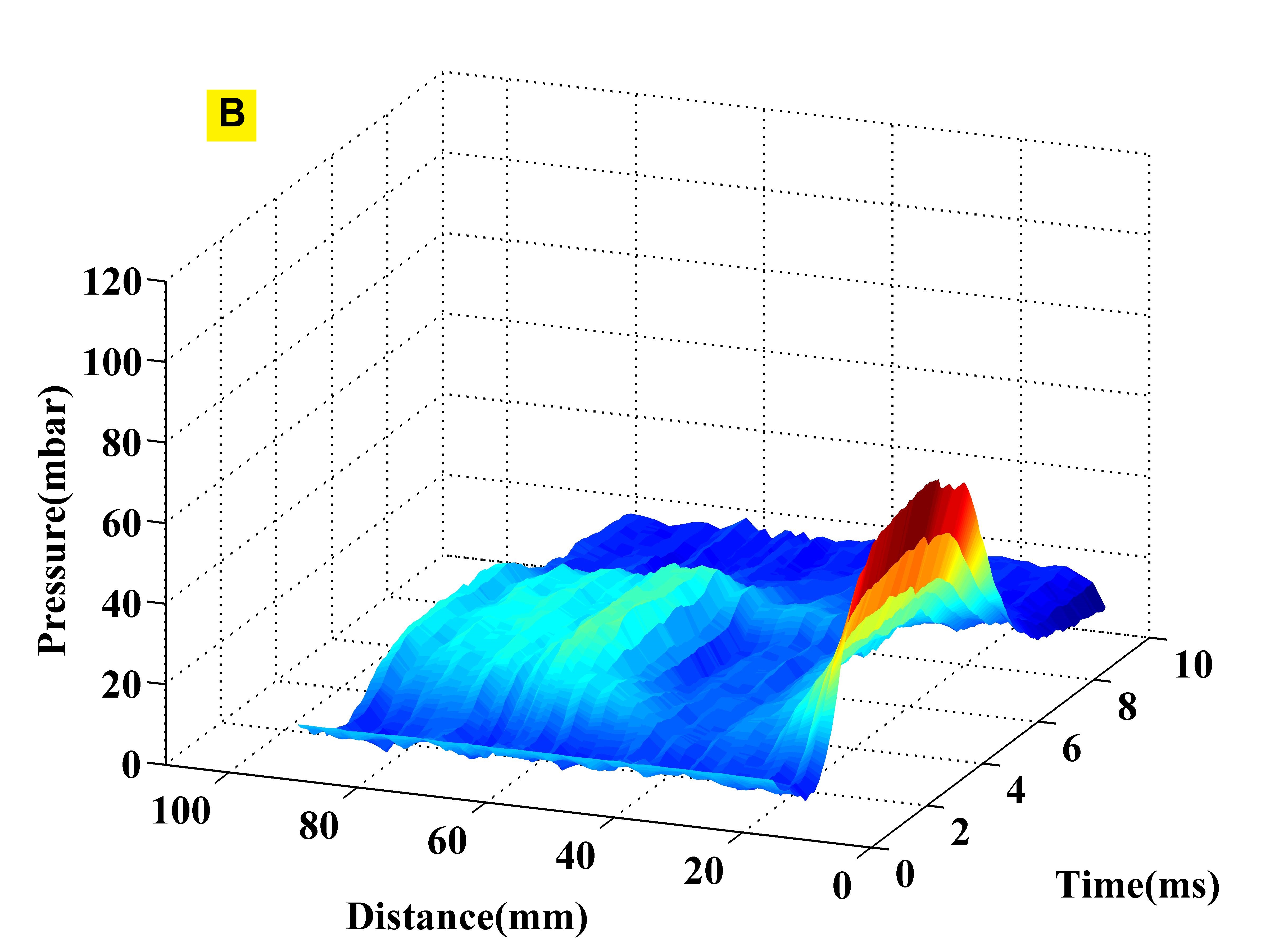}
\caption{\label{fig:PDT_N2_40-2} PDT curves for helium jet[A] and nitrogen jet[B] $P_{0}$:40 bar $P_b$:3mbar. The X, Y and Z axis represents time (ms), distance (D)(mm), and pressure (mbar) respectively.}
\end{figure}

\begin{figure}[ht]
\centering
\includegraphics[width=0.9\columnwidth, trim=0cm 0cm 0cm 0cm, clip=true,angle=0]{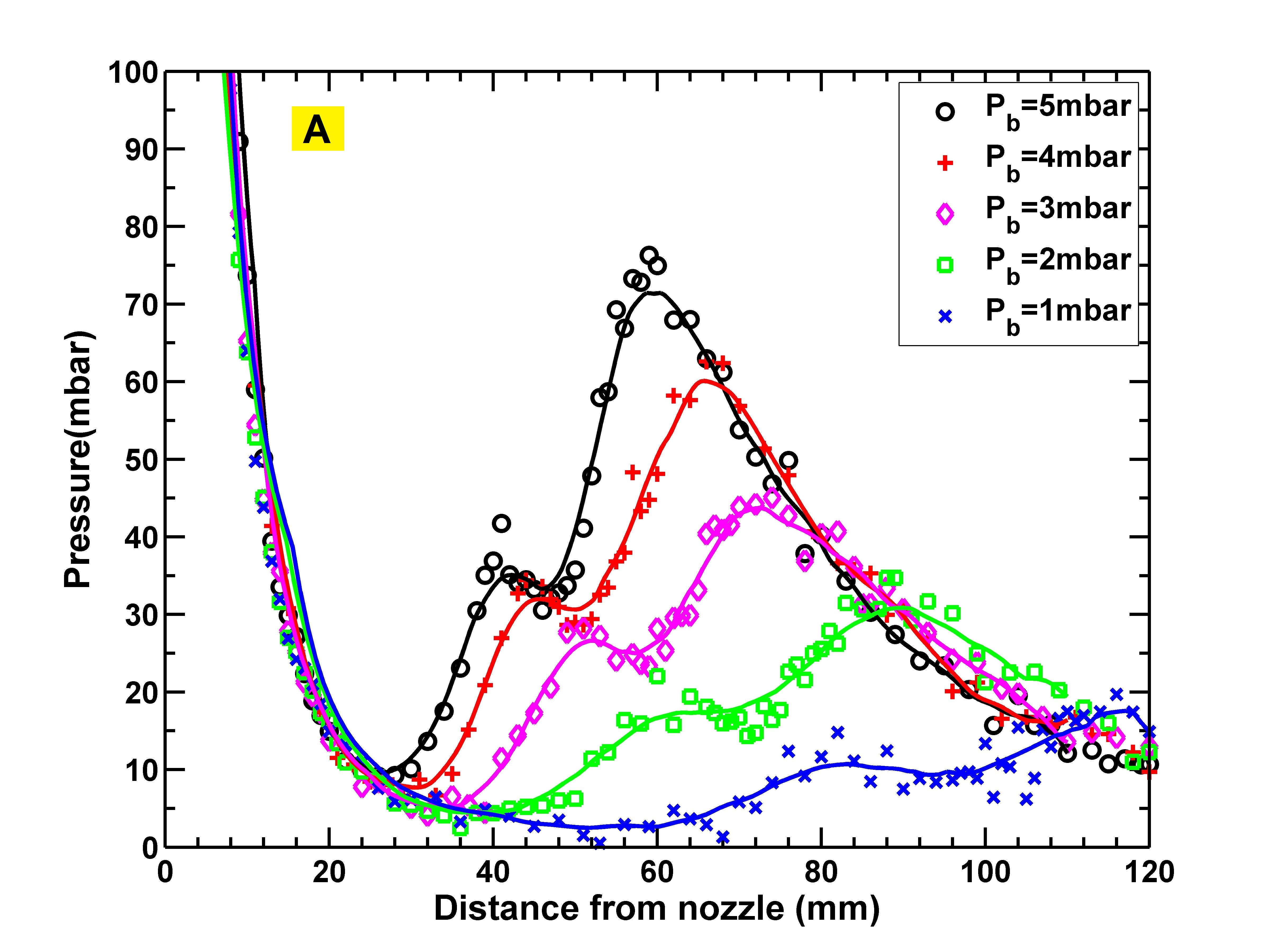}
\includegraphics[width=0.9\columnwidth, trim=0cm 0cm 0cm 0cm, clip=true,angle=0]{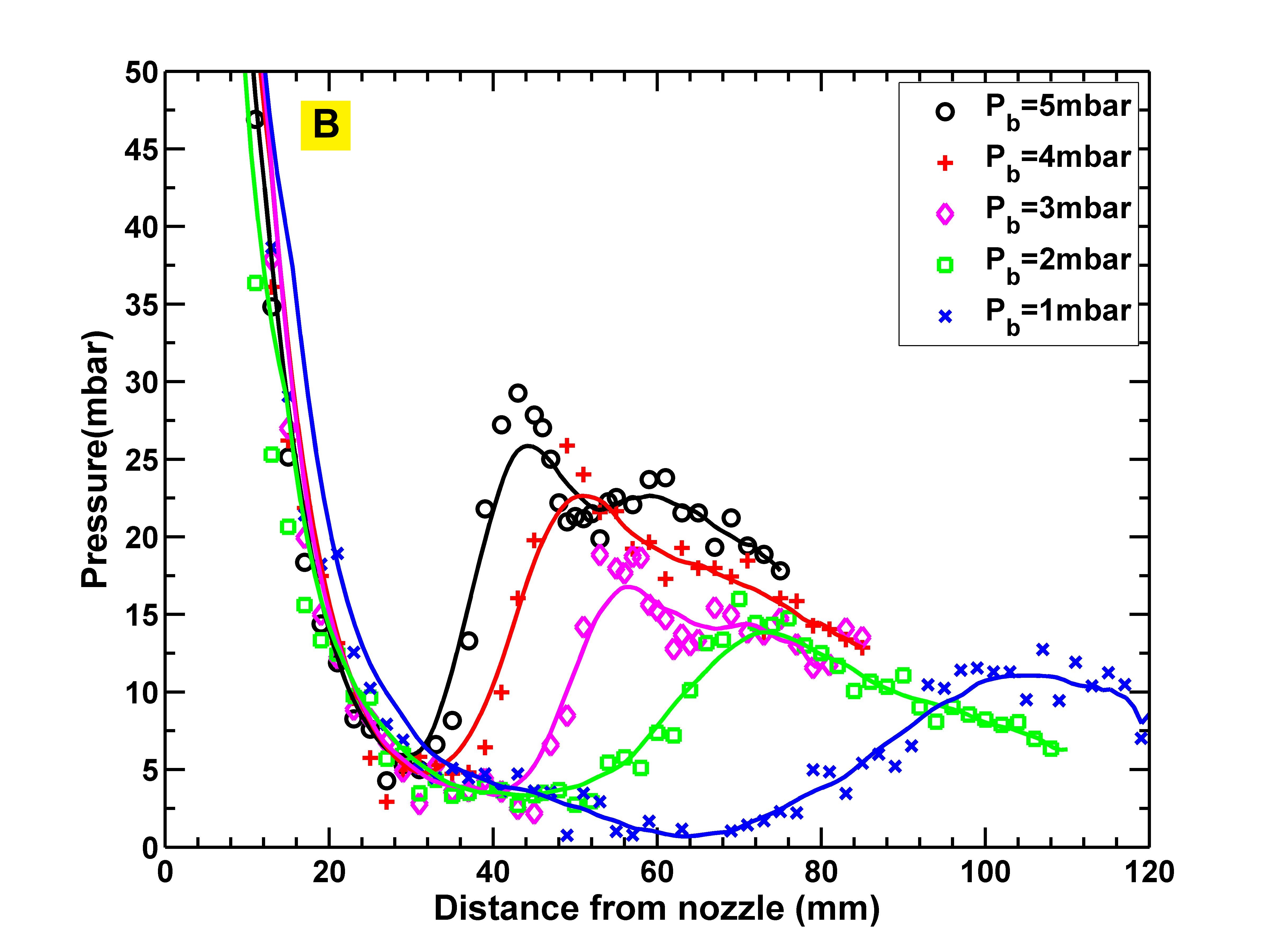}
\caption{\label{fig:pd_he_n2}  Axial pressure distribution for $P_0$ = 40 bar and for 5 different background pressure (table-\ref{table:trans}: set-1) for helium jet[A] and nitrogen jet[B]. The fitted lines are cubic splines.}
\end{figure}

The time evolution of pressure within the jet at various axial distances for 5ms pulse is measured for both nitrogen and helium jets. The information is plotted as a pressure distance time (PDT) curve. The PDT curves for helium and nitrogen jets at $P_0$=40 bar $P_b$=3 mbar respectively are shown in figure-\ref{fig:PDT_N2_40-2}. A rapid fall in pressure is observed within a few exit diameters of the nozzle which indicates that a large part of the expansion occurs within it. Furthermore, the minimum pressure inside the jet is equal to the background pressure. A significant rise in the pressure observed far downstream of the flow indicates the presence of normal shock. The location of normal shock remains nearly unchanged for the entire gas injection duration which again confirms that background pressure rise is insignificant. PDT curve also shows that the output jet pulse shape follows the input pulse. As mentioned earlier, the location of normal shock indicates the boundary of ZOS. Hence, ZOS is taken as the distance from the pulse valve to the location where peak pressure is observed. Pressure vs Distance (PD) curve is obtained by averaging the PDT curves over the pulse duration (2ms to 6ms), which is used to measure the location of peak pressure and hence the length of ZOS. The time-averaged PD curves for data set-1 (table-\ref{table:trans}) are shown in figure-\ref{fig:pd_he_n2}.

 PD curve also shows that with decrease in the background pressure the peaks are shifted away from the pulse valve indicating increase in ZOS size. This is due to decrease in pressure ratio caused by decrease in background pressure. The decrease in the amplitude and broadening of peak pressure at low background pressures indicates that the normal shock gets weakened and diffused. We believe that it could be due to the spread of normal shock along the radial direction, which indicates radial expansion of ZOS thereby increasing the divergence of the jet. However, with flow confined within this chamber of small aspect ratio ($length \approx 4.4\times diameter$), specific comment on the radial size of ZOS can not be made. For significantly high pressure ratios, normal shock will be extremely diffused and may eventually disappear. Of course, a jet in such a case will behave like free expanding gas.

\FloatBarrier

\FloatBarrier

\begin{figure}[ht]
\centering
\includegraphics[width=0.8\columnwidth, trim=0cm 0cm 0cm 0cm, clip=true,angle=0]{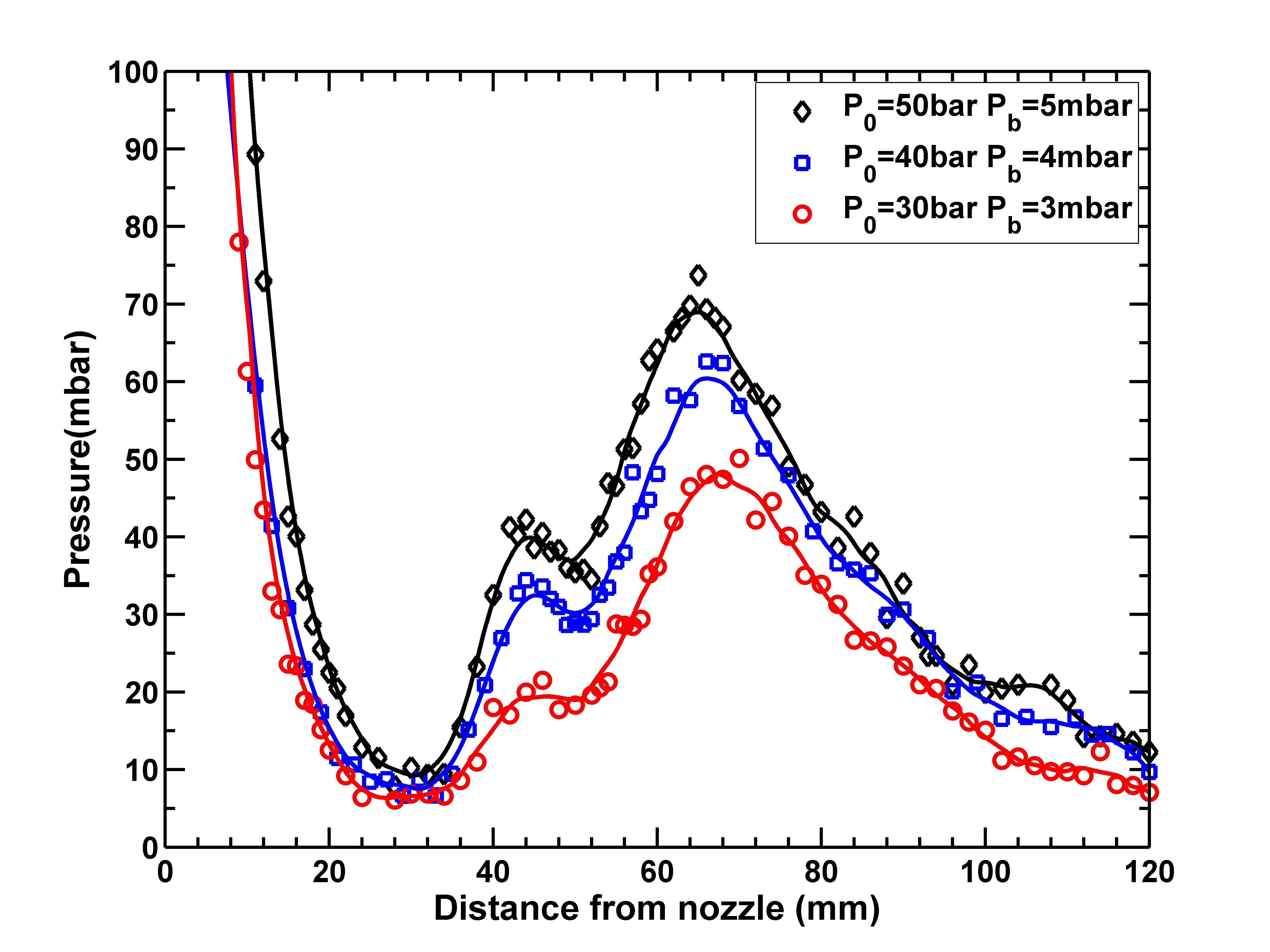}
\caption{\label{fig:set_2} Axial pressure distribution (PD) curves for constant $P_0/P_b$ = $1\times 10^4$ for helium jet. The fitted lines are cubic splines.}
\end{figure}

The measurements of set 2 was carried out for 3 different combinations of $P_0$ and $P_b$ (table:-\ref{table:trans}) so that $P_0/P_b$ is same for all cases and are shown in figure-\ref{fig:set_2}. As can be seen from figure-\ref{fig:set_2}, pressure profile and length of ZOS remains the same for all the three sets of measurements indicating that the length of ZOS depends only on the pressure ratio.

\begin{figure}[ht]
\centering
\includegraphics[width=0.8\columnwidth, trim=0cm 0cm 0cm 0cm, clip=true,angle=0]{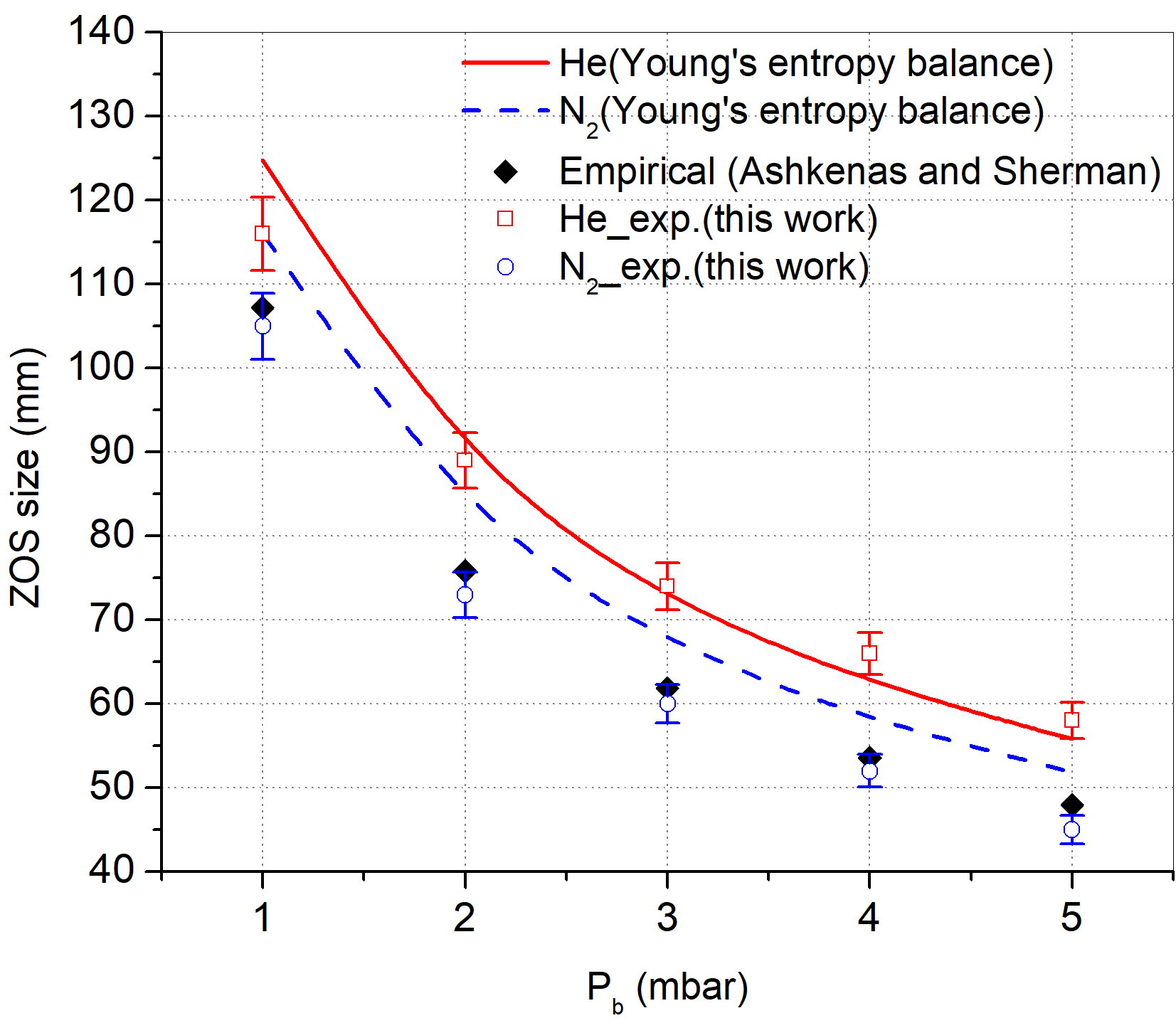}
\caption{\label{fig:ZOS_size} Empirical vs experimental lengths of ZOS ($x_m$) for reservoir pressure ($P_0$) = 40bar and for different background pressures ($P_b$) for helium and nitrogen. The uncertainty in experimental measurement is $\sim$4\%}
\end{figure}

Figure-\ref{fig:ZOS_size} shows a comparison of the estimated lengths of ZOS for helium and nitrogen from this experiment with the empirical relation by Ashkenas and Sherman (equation-\ref{xm}) and Young's entropy balance principle (equation-\ref{xm_g}). Major experimental uncertainties come from the calibration of pressure and gauge used for background pressure measurement. The measurement uncertainty associated with calibration of $P_0$ is 2.5\%, and accuracy of gauge used for background pressure measurement is 5\%. A simple error analysis is used for a cumulative error in $x_m$. It is estimated to be $\sim$4\%. For helium jet, ZOS length remains consistently large compared to that of nitrogen jet as indicated by Young's formula. The experimental ZOS for nitrogen jet appear to closely follow the empirical formula of Ashkenas and Sherman. Even though the empirical relation of ZOS is formulated for Campargue type supersonic source (nozzle sharply cut at sonic plane), a small exit cone acting as a diverging section seem to have no significant effect on the size of ZOS. 

\begin{figure}[ht]
\centering
\includegraphics[width=0.9\columnwidth, trim=0cm 0cm 0cm 0cm, clip= true,angle=0]{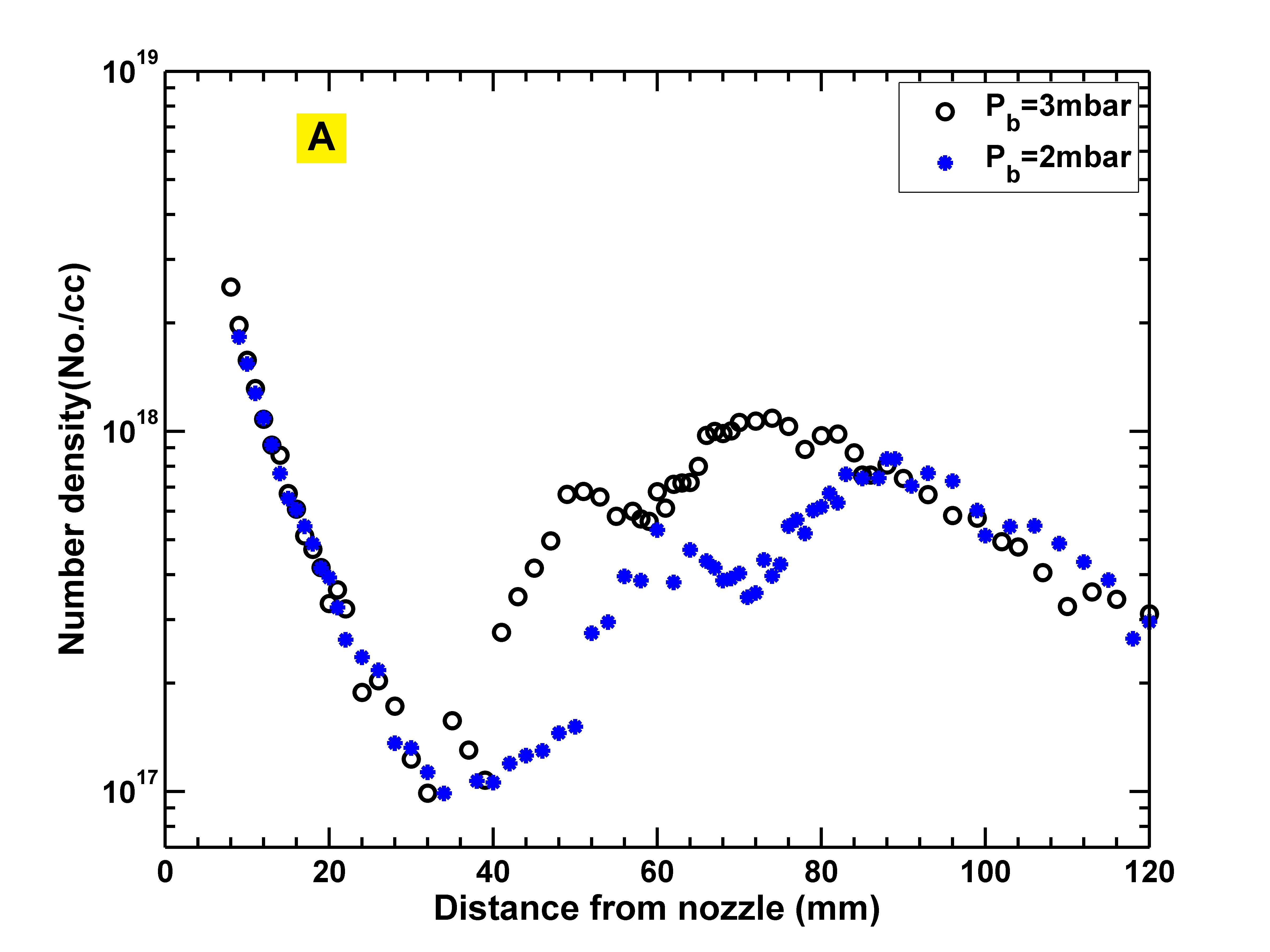}
\includegraphics[width=0.9\columnwidth, trim=0cm 0cm 0cm 0cm, clip= true,angle=0]{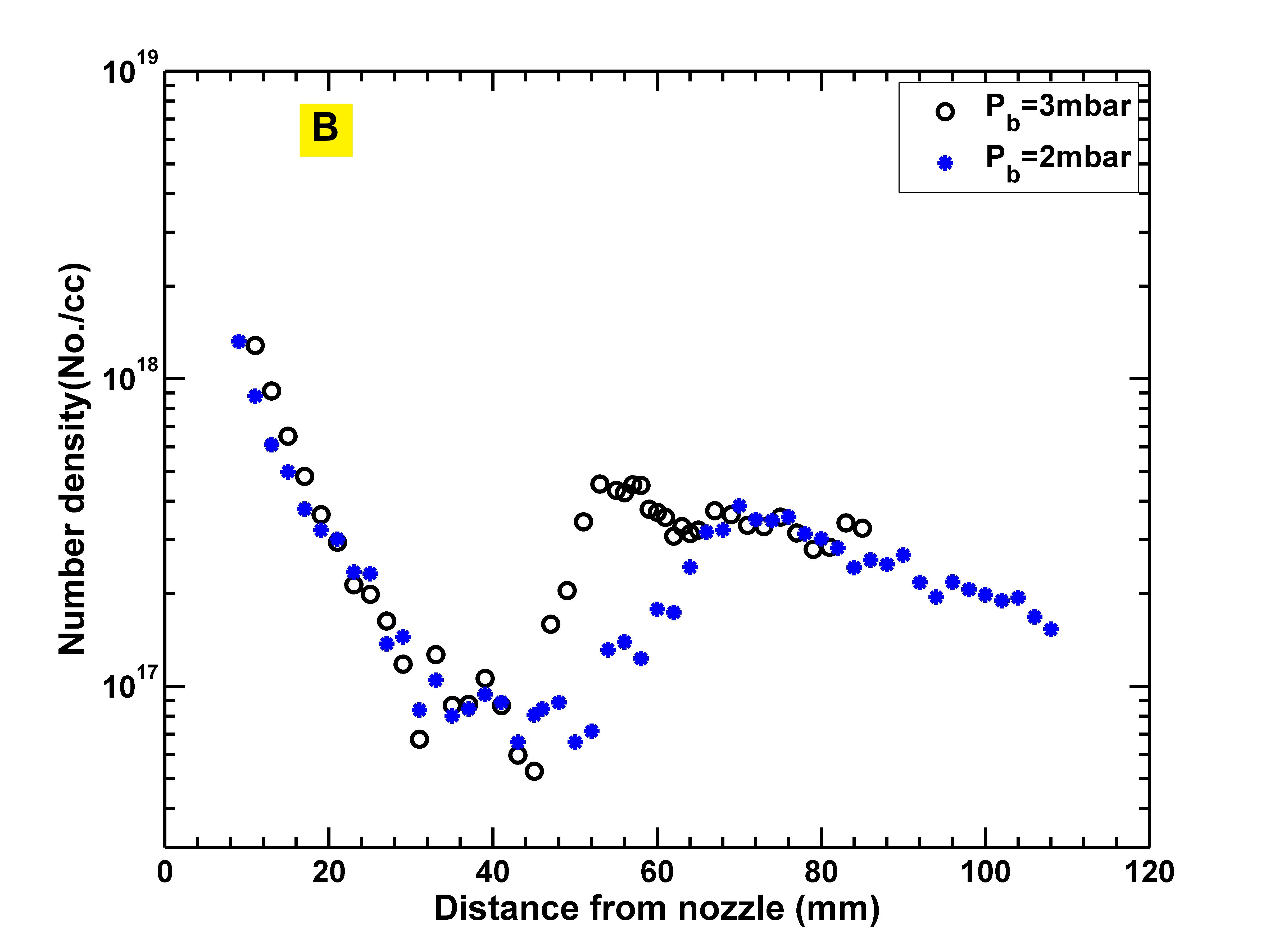}
\caption{\label{fig:dens} Axial Number density plots (No/cc) for helium jet[A] and nitrogen jet [B]: $P_0=40bar$ }
\end{figure}

Number density in the jet is calculated from pressure measured by transducer using ideal gas equation. The temperature of the transducer cavity is considered to be equal to the stagnantion temperature of the reservoir which in the present case is 300K. 1D adiabatic expansion at the jet axis is assumed to be isentropic. Hence, the stagnation temperature would remain equal to reservoir temperature 300K throughout the flow along the axis.  The density plots for helium and nitrogen jets for $P_b$= 40 bar, for two background pressures are shown in figure-\ref{fig:dens}. Density plots follow similar trend as observed by Belan et al.\cite{3_belan2010}. However, no direct comparison with the results of Belan \cite{3_belan2010} can be made due to different kinds of experimental conditions and nozzles.  

\FloatBarrier

\subsection{Estimation of velocity using TOF probe}\label{subsec:result_TOF}

 Velocity measurements for both helium and nitrogen jets are done for $P_0$=40 bar and $P_b$=$5\times10^{-5}$ mbar. As mentioned earlier, measurements are carried out for a pulse duration of 200$\mu s$ for both helium and nitrogen jets. The increase in pressure after injection of helium gas is found to be $\approx 9\times 10^{-5}$ mbar while for nitrogen jet it is $\approx 3\times 10^{-5}$ mbar. The flow rates calculated from source conditions for helium and nitrogen jets are $2.82\times10^{23}$ atoms/sec and $1\times10^{23}$ molecules/sec, respectively. Both experiments and calculation indicates that mass flow rate in helium jet is about 3 times more than in nitrogen jet. This is also observed during transducer experiments (table-\ref{table:trans}).

\begin{figure}[ht]
\centering
\includegraphics[width=0.8\columnwidth, trim=0cm 0cm 0cm 0cm, clip=true,angle=0]{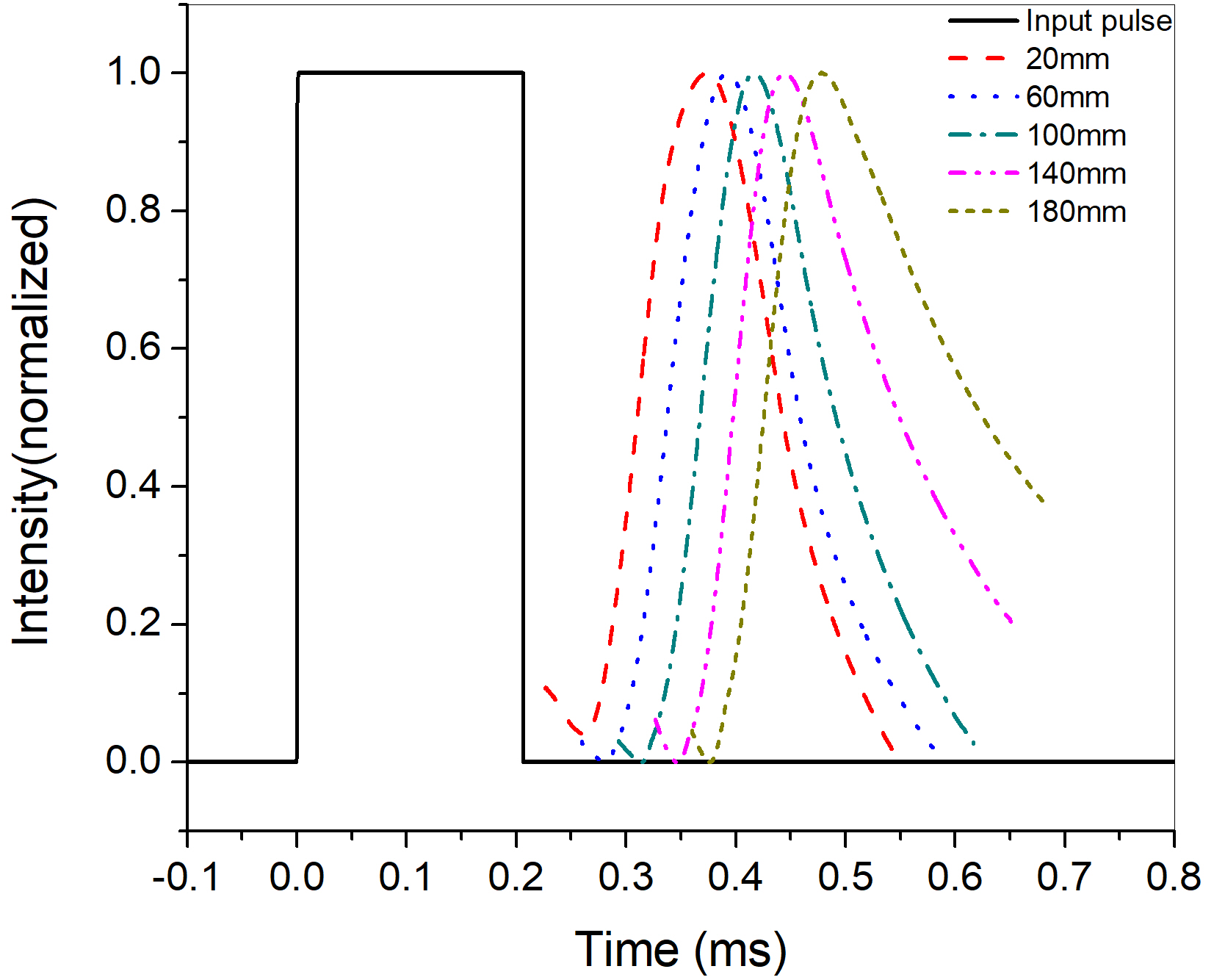}
\caption{\label{fig:tof1_th} Normalized time history plots for TOF probe for Helium jet at different axial distances. (only few plots are shown for the clarity).}
\end{figure}
 
\begin{figure}[ht]
\centering
\includegraphics[width=0.8\columnwidth, trim=0cm 0cm 0cm 0cm, clip=true,angle=0]{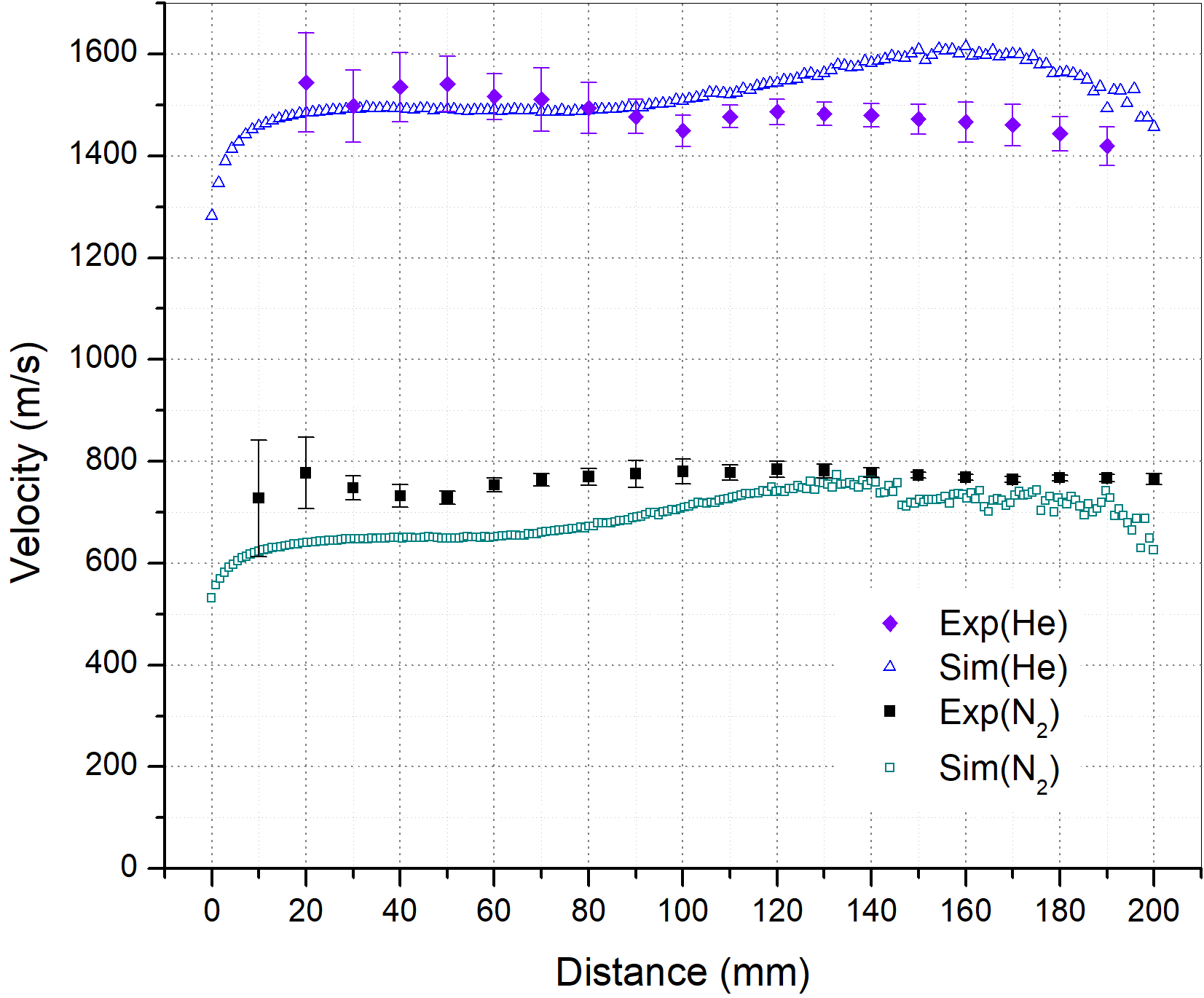}
\caption{\label{fig:dual_probe_result}Comparison of experimentally measured velocities with DSMC simulation for helium and nitrogen jets.}
\end{figure}

For injected gas pulse, change in collector current with time (TOF probe signal) is recorded using a digital oscilloscope. The trigger pulse of the pulse valve is used as an external trigger to the oscilloscope to capture the output signal from the TOF probe. The TOF probe signal is measured for 20 separate distances (D) in steps of 10mm along the axis of the jet up to 200mm from the pulse valve. Here distance (D), is the distance from the collector tip to the exit of the pulse valve. Figure-\ref{fig:tof1_th} shows the normalized time history (TH) plots at various distances from the nozzle. The shift in peak time between two distances is used to determine the velocity of the jet at that location. Each TH plot shown in Figure-\ref{fig:tof1_th} is an average of 5 distinct gas pulses injected at a single location. The TH curves at multiple locations are plotted to determine the velocity profile. Figure-\ref{fig:dual_probe_result} shows the velocity profile generated using 3 sets of experiments with error bars representing standard deviation.

The uncertainty in the measured value of velocity in TOF probe can arise primarily due to jitter of the pulse valve. When using the input TTL as a trigger to the acquisition of TOF probe signal, there is a certain uncertainty involved between the trigger and actual mechanical opening of the valve. This could result in shifting of peak position in TH curve between individual measurements, resulting in statistical errors. Further, it is likely that uncertainties may arise due to ionization and collection times of the probe. However, this is not likely to affect the measurement as only the difference in flight time is used to estimate the velocity. Again, the statistical error is expected to be decreased when averaged over large number of experiments.

Here we would like to mention that we also observe increased uncertainty in velocity for the measurements near the nozzle. This is due to the proximity of the grounded pulse valve which can change the electric field distribution. Moreover, the probe cannot be operated at background pressure $P_b>10^{-4}$ mbar as thermally emitted electrons result in a gas discharge between electrodes creating DC plasma. These thermal electrons compensate for Thomson’s secondary electrons and discharge occurs even at lower pressures. This effect is increased during gas injection and a sudden rise in pressure between electrodes results in gas discharge.

Velocity of Helium measured by TOF probe is 1500 $\pm$ 120m/s which is significantly less compared to the terminal velocity (1760 m/s) in case of ideal expansion (Eqs. \ref{pr_critical} to \ref{v}) through a sonic nozzle without losses. The conical expanding nozzle is optimized for high centerline density \cite{0_kluria_SMBI} and may create characteristic reflections and hence may have non negligible dissipative effects. It is expected that some energy may be lost and thus decrease the final velocity which is expected to be different for mono- and diatomic gases. Also, the collision among the molecules after exiting the nozzle cannot be negligible as conical nozzle is optimized for high center line density compared to sonic nozzle. This also prevents cooling down of gas to achieve close to theoretical terminal velocity. The effect due to the interaction of reflected gas atoms from the end of the vessel may be ruled out in this case as the pulse duration (200 $\mu$s) of the gas jet is small enough (pulse length for helium for terminal velocity comes out to be 352mm) to have a reflected shock from end of the vessel (870mm).  However, the pulse length is still larger than twice the radius of vacuum vessel and hence there is a possibility that reflected atoms form cylindrical surface may interfere with axial flow (depending on divergence of jet) contributing to loss. On the other hand, for nitrogen the measured velocity comes out to be 750 $\pm$ 120 m/s which is, of course closer to the terminal velocity (790 m/s).

\subsection{DSMC simulation}\label{subsec:result_DSMC}
 
\begin{figure}[ht]
\centering
\includegraphics[width=0.8\columnwidth, trim=0cm 0cm 0cm 0cm, clip=true,angle=0]{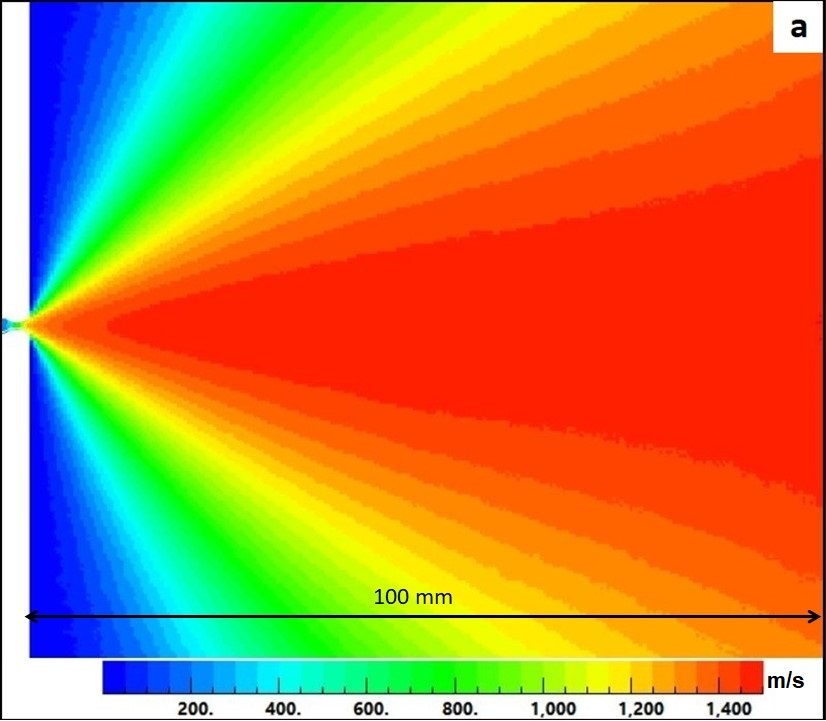}
\includegraphics[width=0.8\columnwidth, trim=0cm 0cm 0cm 0cm, clip=true,angle=0]{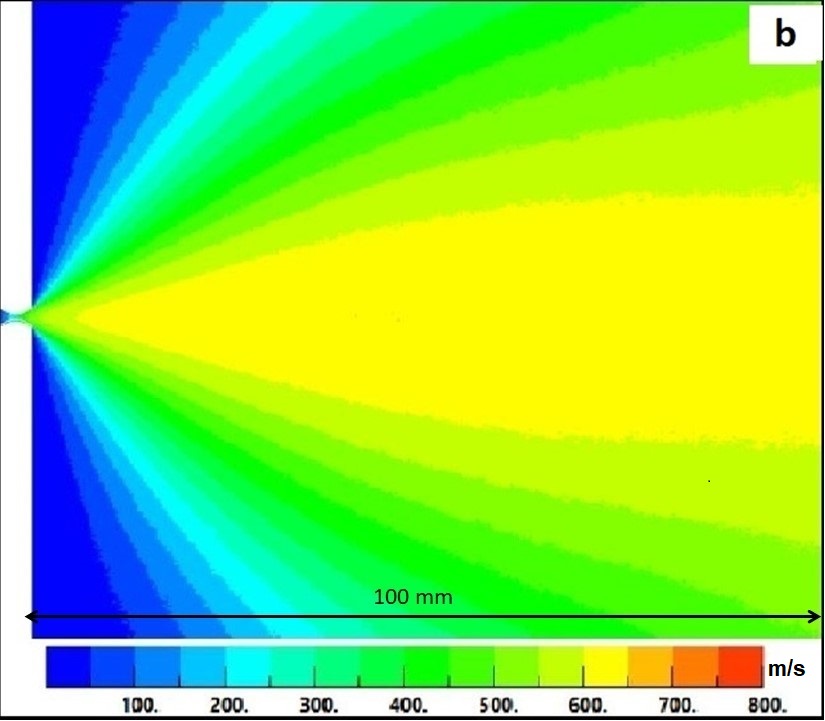}
\caption{\label{fig:DSMCvelocity} Spatial distribution of gas velocity for Helium[a] and Nitrogen[b], from DSMC simulation}
\end{figure}

Figure-\ref{fig:DSMCvelocity} shows the spatial velocity distribution for helium and nitrogen jets simulated using DSMC Simulation. The simulation is carried out for same nozzle geometry as that of the pulse valve described in Section-\ref{sec:theory}. The DSMC method is limited to transition and free molecular flows (Knudsen number $\geq0.1$). In order to maintain the Knudsen number above 0.1, simulation is done at reduced reservoir pressure ($P_0$) of 1mbar, compared to 40 bar in the experiment (Knudsen number is inversely proportional to pressure). Moreover, the TOF experiments could not be done at reservoir pressure ($P_0$) lower than 1 bar because even at $P_0$ = 1 bar, the injected gas is low to be detected by TOF probe for a distance more than 80 mm from the valve. 

 According to equation-\ref{v}, the temperature ($T_0$) of the reservoir is the only reservoir condition affecting the flow velocity. Hence, the spatial velocity distribution is independent of reservoir pressure ($P_0$). This makes it possible to directly compare the simulation results of the velocity with corresponding experimental results as both are at the same reservoir temperature ($T_0$). The dependence of velocity on temperature (and on gamma ratio) refers to the flow around the jet axis so only the axial velocity can be compared safely for experiments and simulations.

\FloatBarrier

Comparison of experimentally measured velocities with the values derived from DSMC simulation is shown in figure-\ref{fig:dual_probe_result}. Velocity profiles of helium and nitrogen appear to be in agreement with their simulated counterparts. However, minor deviations are expected as the adiabatic boundary conditions used in simulations are more ideal than in the experiments. The deviation at the tail is primarily due to proximity of flow boundary. The number density during flow is always greater than the number density corresponding to background pressure applied at the exit boundary which may accelerate the flow close to boundary to accommodate the boundary conditions. This is likely to result in slightly higher than expected velocity observed at the tail of the simulated velocity profile. The simulated velocities for helium also are smaller than the terminal velocity again indicating losses due to nozzle geometry.

\section{Conclusion}\label{sec:conclusion}

Concluding in this work, the pressure and density profiles for helium and nitrogen jets are measured using a pressure transducer.  Length of ZOS for Helium and Nitrogen is estimated  from the measurements. 

The length of ZOS measured is found be in agreement with the established empirical and theoretical models. As expected from the Young's model, ZOS for helium found to be larger than that of nitrogen. We observe that Ashkenas model appears more appropriate for estimation of ZOS for both gases. 

A TOF probe based diagnostics is conceptualised and demonstrated for estimating the velocity profiles of gas jets without obstructing the flow of jet. The measurements for helium jet show that the jet is not fully expanded to its terminal velocity which is also observed in the DSMC simulations. At this point we believe that velocity drop may occur primarily because of losses in the conical expanding section of the nozzle. However, we would like to point out that actual quantification of loss mechanisms needs further studies which are beyond the scope of the present work. DSMC simulations show good agreement with the experimental measurements and hence can be used for precise optimisation of skimmer distances from the nozzles for reducing the skimmer interference as well as for better characterization of the beam. In view of this we anticipate that the results of this study will be helpful in the optimization of beam properties as required for the specific purposes of fuelling or diagnostics etc. Further, the TOF probe demonstrated in the present work will be helpful in furthering the understanding of the supersonic jet. 

\section*{Data availability} 

The data that supports the observations of this study are available from the corresponding author upon reasonable request.


\end{document}